\documentstyle[aps,epsf,rotate,preprint]{revtex}

\begin{document}

\draft

\title{The statistical properties of the volatility of price fluctuations }

\author{Yanhui Liu$^{1}$, Parameswaran Gopikrishnan$^{1}$, Pierre Cizeau$^{1}$, \\
       Martin Meyer$^{1}$, Chung-Kang Peng$^{1,2}$, and H. Eugene Stanley$^{1}$
}

\address{ $^{1}$ Center for Polymer Studies and Dept. of Physics, Boston
        University, Boston, MA 02215, USA \\ $^{2}$ Margret \& H.A. Rey
        Laboratory for Nonlinear Dynamics in Medicine, Beth Israel
        Deaconess Medical Center, Harvard Medical School, Boston 02215 }

\date{Last modified: January 24, 1999.  Printed: \today}

\maketitle

\begin{abstract}

We study the statistical properties of volatility---a measure of how
much the market is likely to fluctuate. We estimate the volatility by
the local average of the absolute price changes. We analyze (a) the S\&P
500 stock index for the 13-year period Jan 1984 to Dec 1996 and (b) the
market capitalizations of the largest 500 companies registered in the
Trades and Quotes data base, documenting all trades for all the
securities listed in the three major stock exchanges in the US for the
2-year period Jan 1994 to Dec 1995. For the S\&P 500 index, the
probability density function of the volatility can be fit with a
log-normal form in the center. However, the asymptotic behavior is
better described by a power-law distribution characterized by an
exponent $1 + \mu \approx 4$. For individual companies, we find a
similar power law asymptotic behavior of the probability distribution of
volatility with exponent $1+\mu \approx 4$.  In addition, we find that
the volatility distribution scales for a range of time
intervals. Further, we study the correlation function of the volatility
and find power law decay with long persistence for the S\&P 500 index
and the individual companies with a crossover at approximately $1.5$
days. To quantify the power-law correlations, we apply power spectrum
analysis and a recently-developed modified root-mean-square analysis,
termed detrended fluctuation analysis (DFA).  For the S\&P 500 stock
index, DFA estimates for the exponents characterizing the power law
correlations are $\alpha_1=0.66$ for short time scales (within $\approx
1.5\,$days) and $\alpha_2=0.93$ for longer time scales (up to a
year). For individual companies, we find $\alpha_1=0.60$ and
$\alpha_2=0.74$, respectively. The power spectrum gives consistent
estimates of the two power-law exponents.

\end{abstract}


\section{Introduction} 

Physicists are increasingly interested in economic time series analysis
for several reasons, among which are the following: ({\it i}) Economic
time series, such as stock market indices or currency exchange rates,
depend on the evolution of a large number of interacting systems, and so
is an example of complex evolving systems widely studied in
physics. ({\it ii}) The recent availability of large amounts of data
allows the study of economic time series with a high accuracy on a wide
range of time scales varying from $\approx 1$ minute up to $\approx
1\,$year. Consequently, a large number of methods developed in
statistical physics have been applied to characterize the time evolution
of stock prices and foreign exchange
rates~\cite{Potters,Mandelbrot63,Man91,Takayasu92x,Zhang,Ausloos,Bouchaud,
Sornette,dietrich,Lux,Solomon,Ghashghaie,Mantegna95}.

Previous
studies~\cite{Potters,Mandelbrot63,Man91,Takayasu92x,Zhang,Ausloos,Bouchaud,
Sornette,dietrich,Lux,Solomon,Ghashghaie,Mantegna95,Pagan96,
Ding93,Dacorogna93,volatility,Cont97,Liu97,pierre,pasquini98,Bak,
Krugman,Gopi98} show that the stochastic process underlying price changes is
characterized by several features.  The distribution of price changes has
pronounced
tails~\cite{Potters,Mandelbrot63,Man91,Lux,Solomon,Ghashghaie,Mantegna95,
  Pagan96} in contrast to a Gaussian distribution. The autocorrelation
function of price changes decays exponentially with a characteristic time of
approximately $4\,$min. However, recent
studies~\cite{Pagan96,Ding93,Dacorogna93,volatility,Cont97,Liu97,pierre,pasquini98}
show that the amplitude of price changes, measured by the absolute value or
the square, shows power law correlations with long-range persistence up to
several months. These long-range dependencies are better modeled by defining
a ``subsidiary process''~\cite{Pagan96,Ding93,Dacorogna93,volatility}, often
referred to as the {\it volatility} in economic literature. The volatility of
stock price changes is a measure of how much the market is liable to
fluctuate. The first step is to construct an estimator for the volatility.
Here, we estimate the volatility as the local average of the absolute price
changes.

Understanding the statistical properties of the volatility also has
important practical implications.  Volatility is of interest to traders
because it quantifies the risk~\cite{Potters} and is the key input of
virtually all option pricing models, including the classic Black and
Scholes model and the Cox, Ross, and Rubinstein binomial models that are
based on estimates of the asset's volatility over the remaining life of
the option~\cite{bs,crr}.  Without an efficient volatility estimate, it
would be difficult for traders to identify situations in which options
appear to be under-priced or overpriced.

We focus on two basic statistical properties of the volatility---the
probability distribution function and the two-point autocorrelation
function. The paper is organized as follows. In Section 2, we briefly
describe the databases used in this study, the S\&P 500 stock index and
individual company stock prices. In Section 3, we discuss the
quantification of volatility. In Section 4, the probability distribution
function is studied, and in Section 5, the volatility correlations are
studied. The appendix briefly describes a recently-developed method,
called detrended fluctuation analysis (DFA) that we use to quantify
power-law correlations.

\section{Data analyzed}

\subsection{S\&P 500 stock index}

The S\&P 500 index from the New York Stock Exchange (NYSE) consists of
500 companies chosen for their market size, liquidity, and industry
group representation in the US. It is a market-value weighted index,
i.e., each stock is weighted proportional to its stock price times
number of shares outstanding.  The S\&P 500 index is one of the most
widely used benchmarks of U.S. equity performance. We analyze the S\&P
500 historical data, for the 13-year period Jan 1984 to Dec 1996
(Fig.~\ref{index-gt}(a)) with a recording frequency of $15$ seconds
intervals. The total number of data points in this 13-year period exceed
4.5 million, and allows for a detailed statistical analysis.

\subsection{Individual company stocks}

We also analyze the Trades and Quotes (TAQ) database which documents
every trade for all the securities listed in the three major US stock
markets---the New York Stock Exchange (NYSE), the American Stock
Exchange (AMEX), and the National Association of Securities Dealers
Automated Quotation (NASDAQ)---for the 2-year period from Jan.~1994 to
Dec.~1995~\cite{TAQ}.  We study the market capitalizations~\cite{Gopi99}
for the 500 largest companies, ranked according to the market
capitalization on Jan.~1 1994. The S\&P500 index at anytime is
approximately the sum of market capitalizations of these 500
companies\cite{spdiff}. The total number of data points analyzed exceed
20~million.

\section{Quantifying Volatility}

The term volatility represents a generic measure of the magnitude of
market fluctuations. Thus, many different quantitative definitions of
volatility are use in the literature. In this study, we focus on one of
these measures by estimating the volatility as the local average of
absolute price changes over a suitable time interval $T$, which is an
adjustable parameter of our estimate.

\begin{figure} 
\centerline{\epsfysize=0.7\columnwidth{{\epsfbox{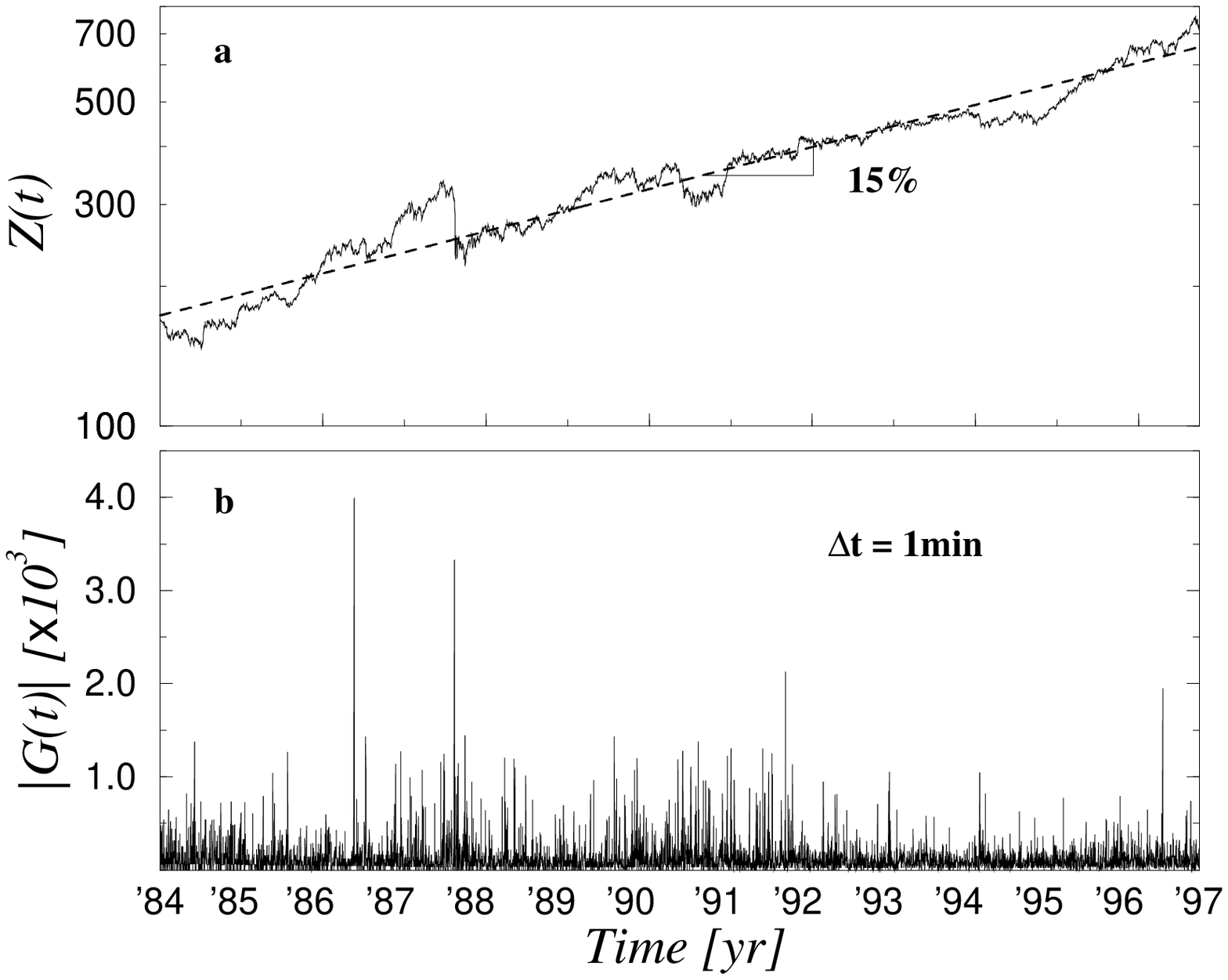}}}}
\vspace*{0.2cm}
\caption{ (a) Data analyzed: The S\&P 500 index $Z(t)$ for the 13-year
period 3 Jan 1984 -- 31 Dec 1996 at sampling intervals $\Delta t =$~1
min. These data include the data set analyzed by Mantegna and
Stanley~\protect\cite{Mantegna95} and the extension of 7 extra
years. Note the large fluctuations, such as that on 19 Oct 1987 (``black
Monday''). The index $Z(t)$ has an increasing trend except for some
crashes, such as the crashes in Oct.~1987 and May 1990. For the period
studied the index can apparently be fit by a straight line on a semi-log
graph, i.e., exponential growth with annual increase rate of $\approx
15\%$. (b) Amplitude of fluctuations, $|G(t)|$ (see text for
definition), with $\Delta t=1\,$min.}
\label{index-gt}
\end{figure}

Fig.~\ref{index-gt}(a) shows the S\&P 500 index $Z(t)$ from 1984 to 1996
in semi-log scale. We define the price change $G(t)$ as the change in
the logarithm of the index,
\begin{equation}
G(t)\equiv\ln Z(t+\Delta t) -\ln Z(t) \cong \frac{Z(t+\Delta t)-Z(t)}{Z(t)}\;,
\label{eq:gt}
\end{equation}
where $\Delta t$ is the sampling time interval. In the limit of small
changes in $Z(t)$ is approximately the relative change, defined by the
second equality. We only count time during opening hours of the stock
market, and remove the nights, weekends and holidays from the data set,
i.e., the closing and the next opening of the market is considered to be
continuous.

The absolute value of $G(t)$ describes the amplitude of the fluctuation,
as shown in Fig.~\ref{index-gt}(b). In comparison to
Fig.~\ref{index-gt}(a), Fig.~\ref{index-gt}(b) does not show visible
global trends due to the logarithmic difference. The large values of
$|G(t)|$ correspond to the crashes and big rallies.

We define the volatility as the average of $|G(t)|$ over a time window
$T=n\cdot\Delta t$, i.e.,

\begin{equation}
V_T(t)\equiv{1\over n}\sum_{t'=t}^{t+n-1}|G(t')|\;,
\label{defV}
\end{equation}
where $n$ is an integer. The above definition can be
generalized~\cite{pasquini98} by replacing $|G(t)|$ with $|G(t)|^\gamma$,
where $\gamma > 1$ gives more weight to the large values of $|G(t)|$ and $0 <
\gamma < 1$ weights the small values of $|G(t)|$.

There are two parameters in this definition of volatility, $\Delta t$
and $n$. The parameter $\Delta t$ is the sampling time interval for the
data and the parameter $n$ is the moving average window size. Note that
the definition of the volatility has an intrinsic error associated with
it. In principle, the larger the choice of time interval $T$, the more
accurate the volatility estimation. However, a large value of $T$ also
implies poor resolution in time.

\begin{figure}
\centerline{
\epsfysize=0.4\columnwidth{{\epsfbox{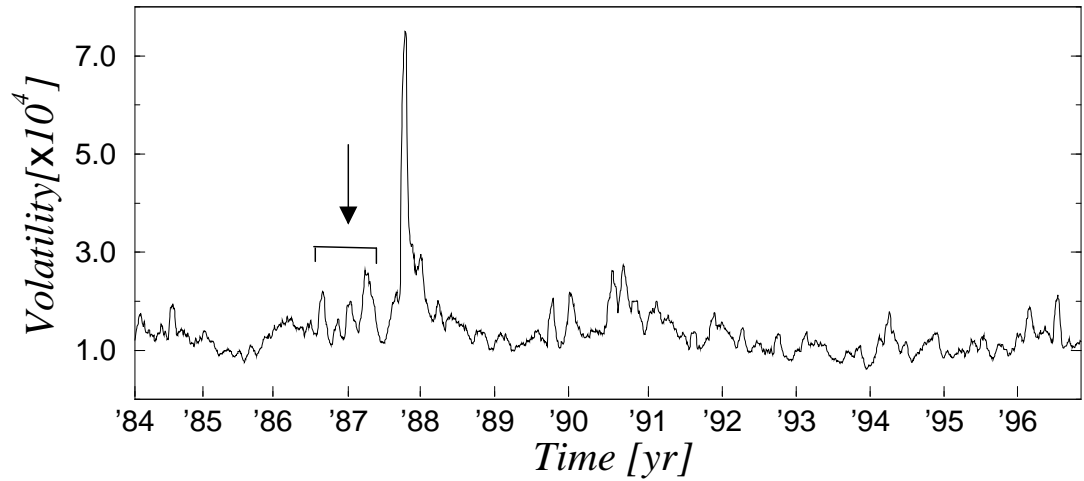}}}
}
\caption{Volatility $V_{T}(t)$ with $T=1\,$month (8190 min) and sampling
time interval $\Delta t = 30\,$min of the S\&P 500 index for the entire
13-year period 1984-96. The highlighted block shows possible
``precursors'' of the Oct.~'87 crash.}
\label{vol_def}
\end{figure}

Fig.~\ref{vol_def} shows the calculated volatility $V_T(t)$ for a large
averaging window $T=8190\,$min (about 1$\,$month) with $\Delta t =
30\,$min. The volatility fluctuates strongly during the crash of '87. We
also note that periods of high volatility are not sparse but tend to
``cluster''. This clustering is especially marked around the '87 crash.
The oscillatory patterns before the crash could be possible precursors
(possibly related to the oscillatory patterns postulated
in~\cite{Bouchaud,Sornette}). Clustering also occurs in other periods,
e.g.~in the second half of '90. There are also extended periods where
the volatility remains at a rather low level, e.g.~the years of '85 and
'93.

\section{Volatility distribution}

\subsection{Volatility distribution of the S\&P 500 index}

\subsubsection{Center part of the distribution}

\begin{figure}
\centerline{
\epsfysize=0.4\columnwidth{{\epsfbox{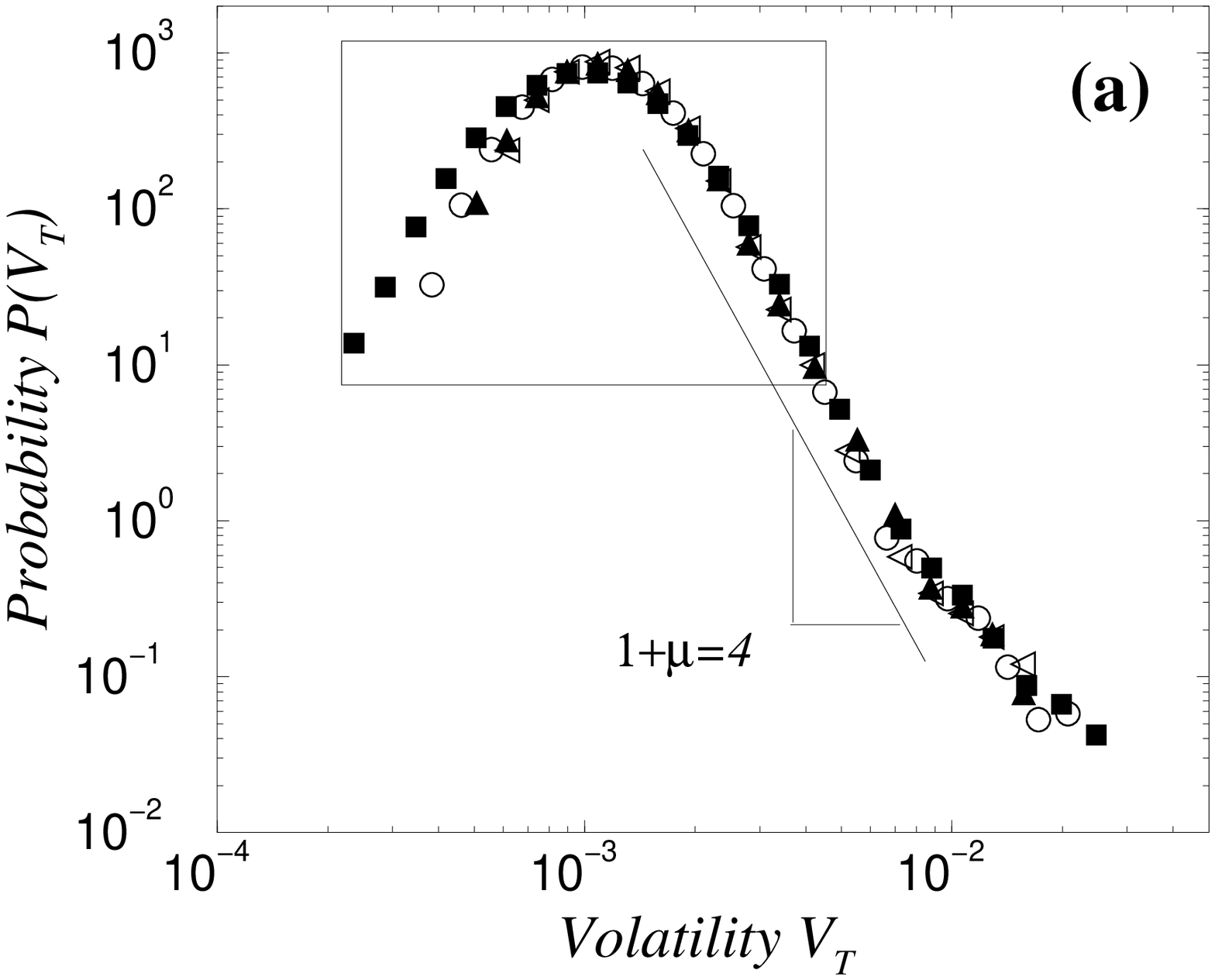}}}
}
\vspace*{2cm}
\centerline{
\epsfysize=0.4\columnwidth{{\epsfbox{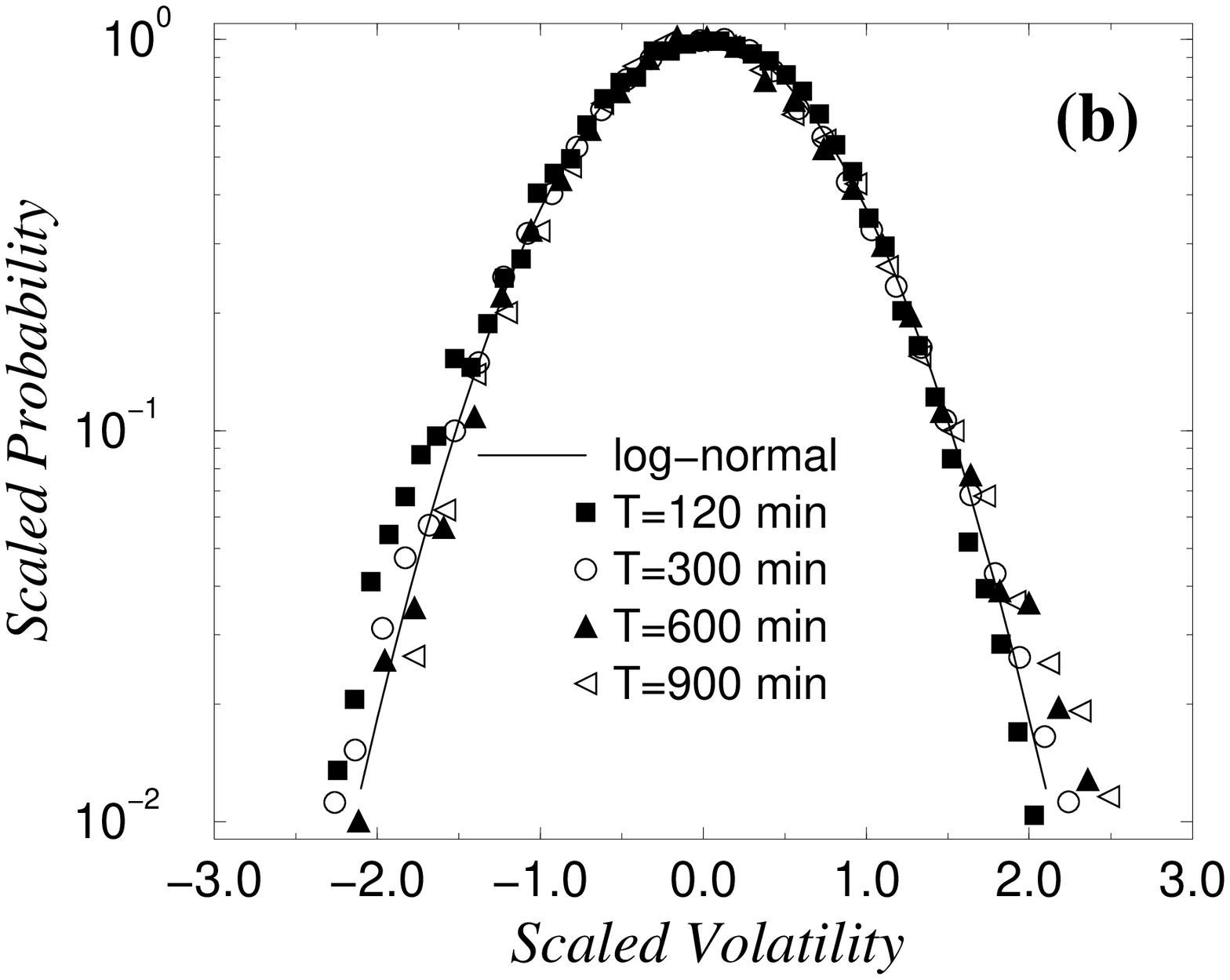}}}
}
\caption{ (a) Probability distribution of the volatility on a log-log
scale with different time windows $T$ with $\Delta t=30\,$min. The
center part of the distribution shows a quadratic behavior on the
log-log scale. The asymptotic behavior seems consistent with a
power-law.  (b) Center of the distribution: The volatility distribution
for different window sizes T using the log-normal scaling form,
$\protect\sqrt{\nu}\exp(a+\nu/4)P(V_{T})$ as a function of
$(\ln(V_{T})-a)/\protect\sqrt{\pi\nu}$, where $a$ and $\nu$ are the mean
and the width on a logarithmic scale. The scaled distributions are shown
for the region shown by the box in (a). By the scaling, all curves
collapse to the log-normal form with $a=0$ and $\nu=1$, $\exp(-(\ln
x)^2)$ (solid line).}
\label{lognormal}
\end{figure}

\begin{figure}
\centerline{
\epsfysize=0.4\columnwidth{{\epsfbox{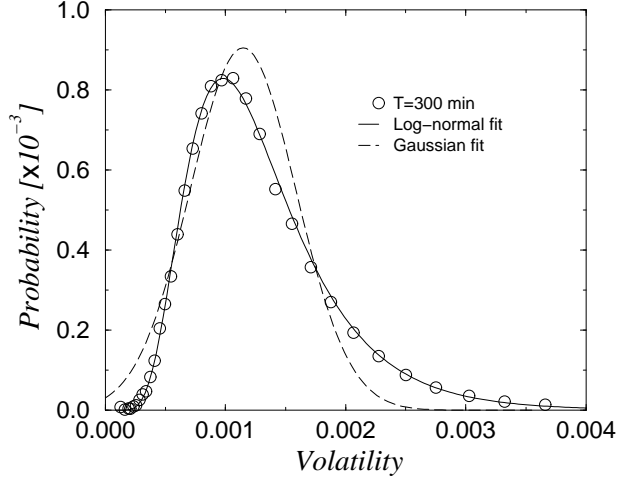}}}
}
\caption{Comparison of the log-normal and Gaussian fits for the
volatility distribution for $T=300\,$min and $\Delta t=30$~min.}

\label{log2}
\end{figure}

Fig.~\ref{lognormal}(a) shows the probability density function $P(V_T)$
of the volatility for several values of $T$ with $\Delta t=30\,$min. The
central part shows a quadratic behavior on a log-log scale
(Fig.~\ref{lognormal}(a)), consistent with a log-normal
distribution~\cite{pierre}. To test this possibility, the
appropriately-scaled distribution of the volatility is plotted on a
log-log plot (Fig.~\ref{lognormal}(b)). The distributions of volatility
$V_T$, for various choices of $T$ (from $T=120$~min up to $T=900$~min),
collapse onto one curve and are well fit in the center by a quadratic
function on a log-log scale. Since the central limit theorem holds also
for correlated series~\cite{Beran94}, with a slower convergence than for
non-correlated processes~\cite{Potters,Solomon,Cont97}, in the limit of
large values of $T$, one expects that $P(V_T)$ becomes
Gaussian. However, a log-normal distribution fits the data better than a
Gaussian, as is evident in Fig.~\ref{log2} which compares the best
log-normal fit with the best Gaussian fit for the data~\cite{pierre}.
The apparent scaling behavior of volatility distribution could be
attributed to the long persistence of its autocorrelation
function~\cite{Cont97} (Section 5).

\subsubsection{Tail of the distribution}

Although the log-normal seems to describe well the center part of the
volatility distribution, Fig.~\ref{lognormal}(a) suggests that the
distribution of the volatility has quite different behavior in the
tail. Since our time window $T$ for estimating volatility is quite
large, it is difficult to obtain significant statistics for the
tail. Recent studies of the distribution for price changes report power
law asymptotic behavior~\cite{Lux,Pagan96,Gopi98}. Since the volatility
is the local average of the absolute price changes, it is possible that
a similar power law asymptotic behavior might characterize the
distribution of the volatility. Hence we reduce the time window $T$ and
focus on the ``tail'' of the volatility.

\begin{figure}
\centerline{
\epsfysize=0.4\columnwidth{{\epsfbox{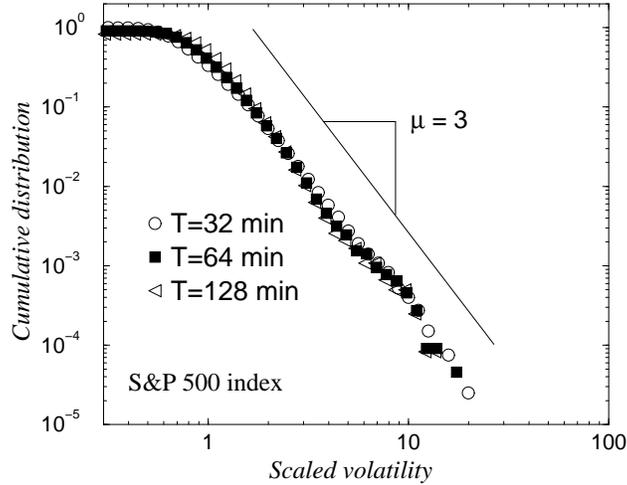}}}
}
\caption{ (a)The cumulative distribution function of the volatility,
scaled by the standard deviation, for time scales $T=32,64,128$ min with
sampling time interval $\Delta = 1\,$min, using non-overlapping windows
for the S\&P 500 stock index. Regression lines yield estimates of the
exponent $\mu = 3.10 \pm 0.08 $ for $T=32\,$min, $\mu = 3.19 \pm 0.10 $
for $T=64\,$min and $\mu = 3.30 \pm 0.15 $ for $T=128\,$min. The fits
were performed over the range of scaled volatility greater than 1
standard deviation. Choices of $\Delta$ from 16~min and 32~min were also
studied for the same values of $T$ shown. Results obtained for these
cases and the values of $\mu$ obtained are consistent with the present
case.}
\label{cum_sp}
\end{figure}

We compute the cumulative distribution of the
volatility. Eq.~(\ref{defV}) for different time scales,
Fig.~\ref{cum_sp}(a). We find that the cumulative distribution of the
volatility is consistent with a power law asymptotic behavior,
\begin{equation}
P(V_T>x) \sim {1\over x^{\mu}}\,.
\label{def_alpha}
\end{equation} 
Regression fits yield estimates $\mu=3.10\pm 0.08$ for $T=32\,$min with
$\Delta t = 1\,$min, well outside the stable L\'evy range $0 < \mu < 2$.

For larger time scales the asymptotic behavior is difficult to estimate
because of poor statistics at the tails.  In view of the power law
asymptotic behavior for the volatility distribution, the drop-off of
$P(V_T)$ for low values of the volatility could be regarded as a
truncation to the power law behavior, as opposed to a log-normal.

\subsection {Volatility distribution for individual companies}

In this section, we extend the investigation of the nature of this
distribution to the individual companies comprising the S\&P 500, where
the amount of data is much larger, which allows for better sampling of
the tails.

From the TAQ data base, we analyze 500 time series $S_i(t)$, where $S_i$
is the market capitalization of company $i$ (i.e., the stock price
multiplied with the number of outstanding shares), $i=1,\dots,500$ is
the rank in descending order of the company according to its market
capitalization on 1 Jan.~1994 and the sampling time is 5~min.The basic
quantity studied for individual stocks is the change in logarithm of the
market capitalization for each company,
\begin{equation}
G_i(t)\equiv \ln S_i(t+\Delta t) -\ln S_i(t) \cong {S_i(t+\Delta
t)-S_i(t)\over S_i(t)}\;,
\label{eq_defG}
\end{equation}
where the $S_i$ denotes the market capitalization of stock $i= 1, \dots
,500$ and $\Delta t=5\,$min.

As before, we estimate the volatility at a given time by averaging
$|G_i(t)|$ over a time window $T=n\cdot\Delta t$,
\begin{equation}
V^i_T \equiv V^i_T(t)\equiv{1\over n}\sum_{t'=t}^{t+n-1}|G_i(t')|\;.
\end{equation}
A normalized volatility is then defined for each company,
\begin{equation}
v^i_T \equiv v^i_T(t)\equiv \frac{V^i_T}{\sqrt{\langle [V^i_T]^2
\rangle - \langle V^i_T \rangle ^2}}\;,
\end{equation}
where $\langle \dots \rangle$ denotes the time average estimated by
non-overlapping windows for different time scales $T$.

\begin{figure}
\centerline{
\epsfysize=0.5\columnwidth{\rotate[r]{\epsfbox{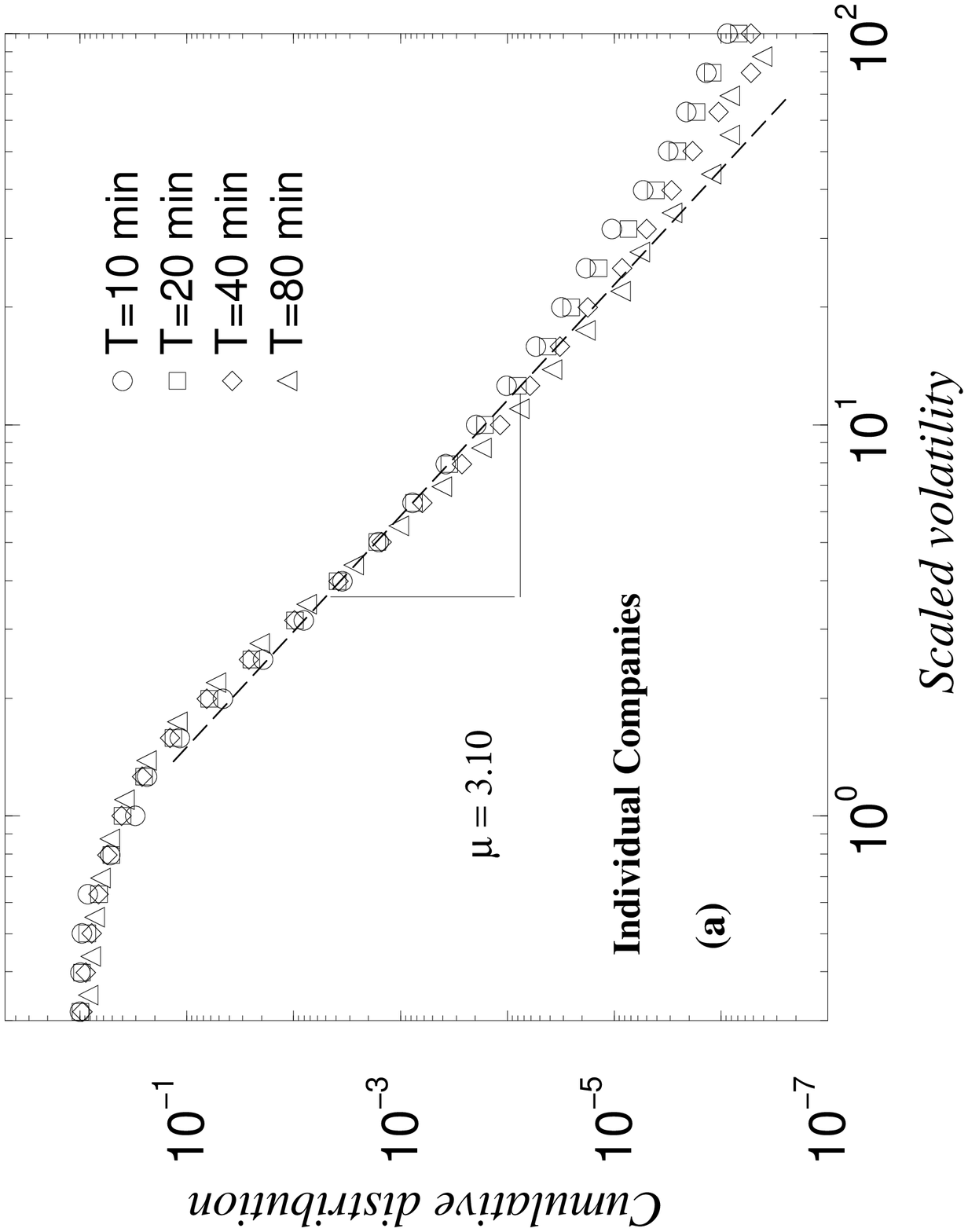}}}
}
\vspace{1cm}
\centerline{
\epsfysize=0.5\columnwidth{\rotate[r]{\epsfbox{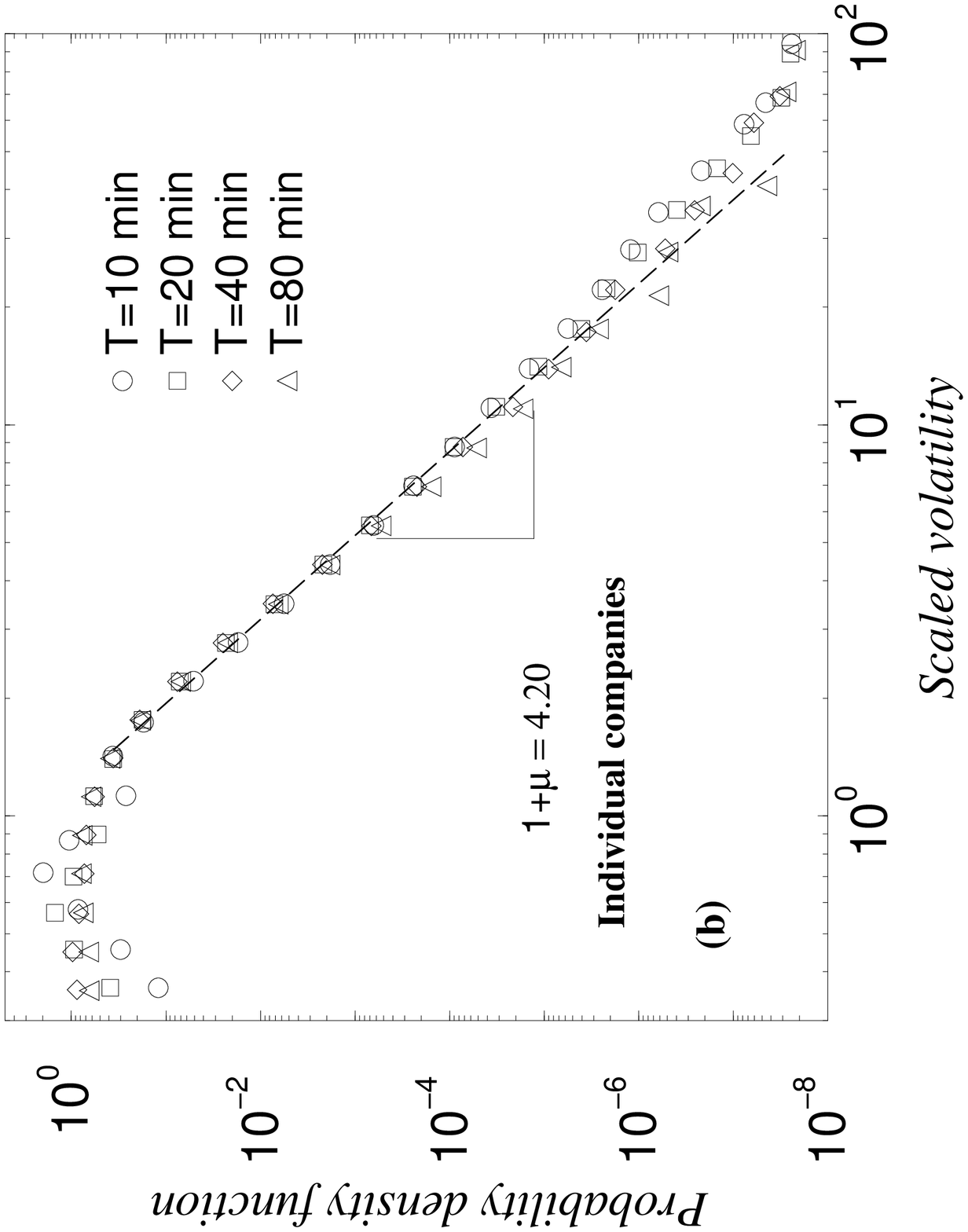}}}
}
\caption{ (a) The cumulative probability distribution on a log-log scale
of the normalized volatility for all the $500$ individual companies for
various averaging window lengths, with a sampling time $\Delta
t=5\,$min. Power law regression fits yield $\mu =3.10 \pm 0.11$ for
$T=10$~min, $\mu =3.16 \pm 0.15$ for $T=20$~min, $\mu =3.28 \pm 0.17$
for $T=40$~min, and $\mu = 3.38 \pm 0.18$ for $T=80$~min. These fits
were performed in the region of scaled volatility between 1 and 30
standard deviations. (b) The probability density function of the
normalized volatility for single companies. Regression fits yield a
slope of $1+\mu = 4.06 \pm 0.10$ for $T=10$~min, $1+\mu = 4.15 \pm 0.13$
for $T=20$~min, $1+\mu = 4.22 \pm 0.15$ for $T=40$~min, and $1+\mu =
4.38 \pm 0.16$ for $T=80$~min. The fits were performed in the region of
scaled volatility between 1 and 50 standard deviations.}
\label{cum_taq}
\end{figure}

Fig.~\ref{cum_taq}(a) shows the cumulative probability distribution of
the normalized volatility $v_T^i$ for all 500 companies with different
averaging windows $T$, where the sampling interval $\Delta t=5\,$min. We
observe a power law behavior,
\begin{equation}
P(v_T^i > x) \sim {1\over x^{\mu}}\,,
\label{pld-cu}
\end{equation} 
Regression fits yield $\mu=3.10 \pm 0.11$ for $T=10$~min. This behavior
is confirmed by the probability density function shown in
Fig.~\ref{cum_taq}(b),
\begin{equation}
P(v_T) \sim {1\over v_T^{\mu + 1}}\,.
\label{pld-de}
\end{equation} 
with a cutoff at small values of the volatility. Regression fits yield
the estimate $1+\mu = 4.06 \pm 0.10$ for $T=10$~min, in good agreement
with the estimate of $\mu$ from the cumulative distribution. Both the
probability density and the cumulative distribution, Figs.~\ref{pld-cu}
and ~\ref{pld-de}, show that the volatility distribution for individual
companies are consistent with power-law asymptotic exponent $\mu \approx
3$, in agreement with the asymptotic behavior of the volatility
distribution for the S\&P 500 index.

In summary, the asymptotic behavior of the cumulative volatility
distribution is well described by a power law behavior with exponent
$\mu \approx 3 $ for the S\&P 500 index. This power law behavior also
holds for individual companies with similar exponent $\mu \approx 3$ for
the cumulative distribution, with a drop-off at low values.

\section{Correlations in the volatility}

\subsection{Volatility correlations for S\&P 500 stock index}

Unlike price changes that are correlated only on very short time
scales~\cite{Fama70}(a few minutes), the absolute values of price changes
show long-range power-law correlations on time scales up to a year or
more~\cite{Pagan96,Ding93,Dacorogna93,volatility,Cont97,Liu97,pierre,pasquini98}.
Previous works have shown that understanding the power-law correlations,
specifically the values of the exponents, can be helpful for guiding the
selection of models and mechanisms~\cite{Bak}.  Therefore, in this part we
focus on the {\it quantification\/} of power-law correlations of the
volatility. To quantify the correlations, we use $|G(t)|$ instead of
$V_T(t)$, i.e. time window $T$ is set to $1\,$min with $\Delta t = 1\,$min
for the best resolution.

\subsubsection{Intra-day pattern removal}

\begin{figure}
\centerline{ \epsfysize=0.5\columnwidth{{\epsfbox{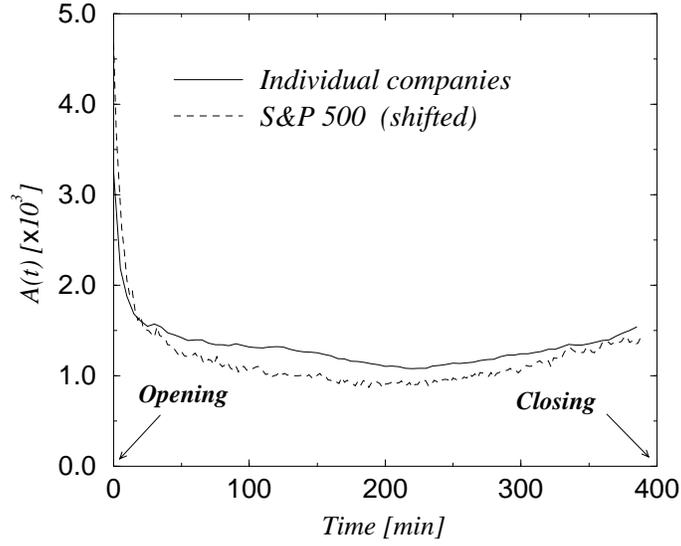}}}
}
\caption{The 1-minute interval intra-day pattern for absolute price
changes of the S\&P 500 stock index (1984-96) (shifted) and for the
absolute price changes, averaged for the chosen 500 companies
(1994-95). The length of the day is 390 minutes. In order to avoid the
detection of spurious correlations, this daily pattern is
removed. Otherwise one finds peaks in the power spectrum at the
frequencies of 1$\,$/day and larger. Note that both the curves have a
similar pattern with large values within the first 15 minutes after the
market opens.}
\label{taq_intraday}
\end{figure}

It is known that there exist intra-day patterns of market activity in
the NYSE and the S\&P 500 index~\cite{Wood,Harris,Admati88,day1}. A
possible explanation is that information gathers during the time of
closure and hence traders are active near the opening hours. And many
liquidity traders are active near the closing hours~\cite{intraday}.
We find a similar intra-day pattern in the absolute price changes
$|G(t)|$ (Fig.~\ref{taq_intraday}). In order to quantify the
correlations in absolute price changes, it is important to remove this
trend, lest there might be spurious correlations. The intra-day pattern
$A(t_{day})$, where $t_{day}$ denotes the time in a day, is defined as the
average of the absolute price change at time $t_{day}$ of the day for
all days.

\begin{equation}
A(t_{day}) \equiv \frac{\sum_{j=1}^{N}|G^{j}(t_{day})|}{N}\;,
\label{eq:at}
\end{equation}
where the index $j$ runs over all the trading days $N$ in the $13$-year
period ($N=3309$ in our study) and $t_{day}$ denotes the time in the
day.  In order to avoid the artificial correlation caused by this daily
oscillation, we remove the intra-day pattern from $G(t)$ which we
schematically write as:

\begin{equation}
g(t)\equiv G(t_{day})/A(t_{day})\;,
\label{eq:norgt}
\end{equation}
for all days. Each data point $g(t)$, denotes the normalized absolute
price change at time $t$, which is computed by dividing each point
$G(t_{day})$ at time $t_{day}$ of the day by $A(t_{day})$ for all days.

Three methods---correlation function, power spectrum and detrended
fluctuation analysis (DFA)--- are employed to quantify the correlation
of the volatility. The pros and cons of each method and the relations
between them are described in the Appendix.

\subsubsection {Correlation quantification}

\begin{figure}
\centerline{
\epsfysize=0.5\columnwidth{{\epsfbox{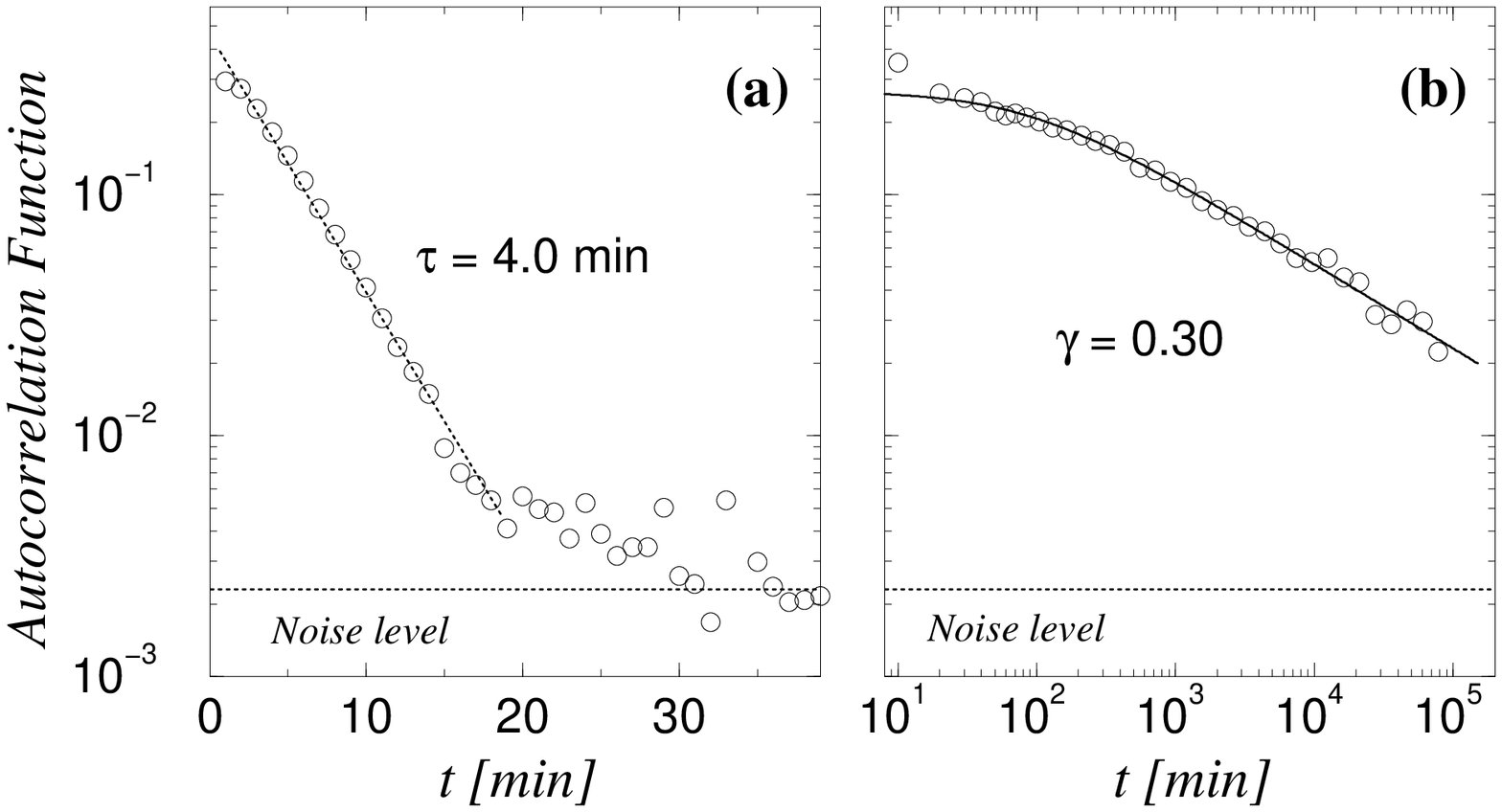}}}
}
\caption{ (a) Semi-log plot of the autocorrelation function of $g(t)$,
(b) Autocorrelation function of $|g(t)|$ in the double log plot, with
sampling time interval $\Delta t=1\,$min. The autocorrelation function
of $g(t)$ decays exponentially to zero within half an hour, $C(t) \sim
\exp (- t/\tau )$ with $\tau \approx 4.0\,$min. A power law correlation
, $C(t) \sim t^{-\gamma}$, exists in the $|g(t)|$ for more than 3
decades. Note that both graphs are truncated at the first zero value of
$C(t)$. The solid line in (b) is the fit to the function ${1 \over
1+t^\gamma}$ from which we obtain $\gamma=0.30 \pm 0.08$. The horizontal
dashed line indicates the noise level.}
\label{correlation}
\end{figure}

Fig.~\ref{correlation}(a) shows the autocorrelation function of the
normalized price changes, $g(t)$, which shows exponential decay with a
characteristic time of the order of 4$\,$min. However, we find that the
autocorrelation function of $|g(t)|$ has power law decay, with long
persistence up to several months, Fig.~\ref{correlation}(b). This result
is consistent with previous studies on several economic time
series~\cite{Pagan96,Ding93,Dacorogna93,volatility,Cont97,Fama70}.

\begin{figure}
\centerline{
\epsfysize=0.4\columnwidth{{\epsfbox{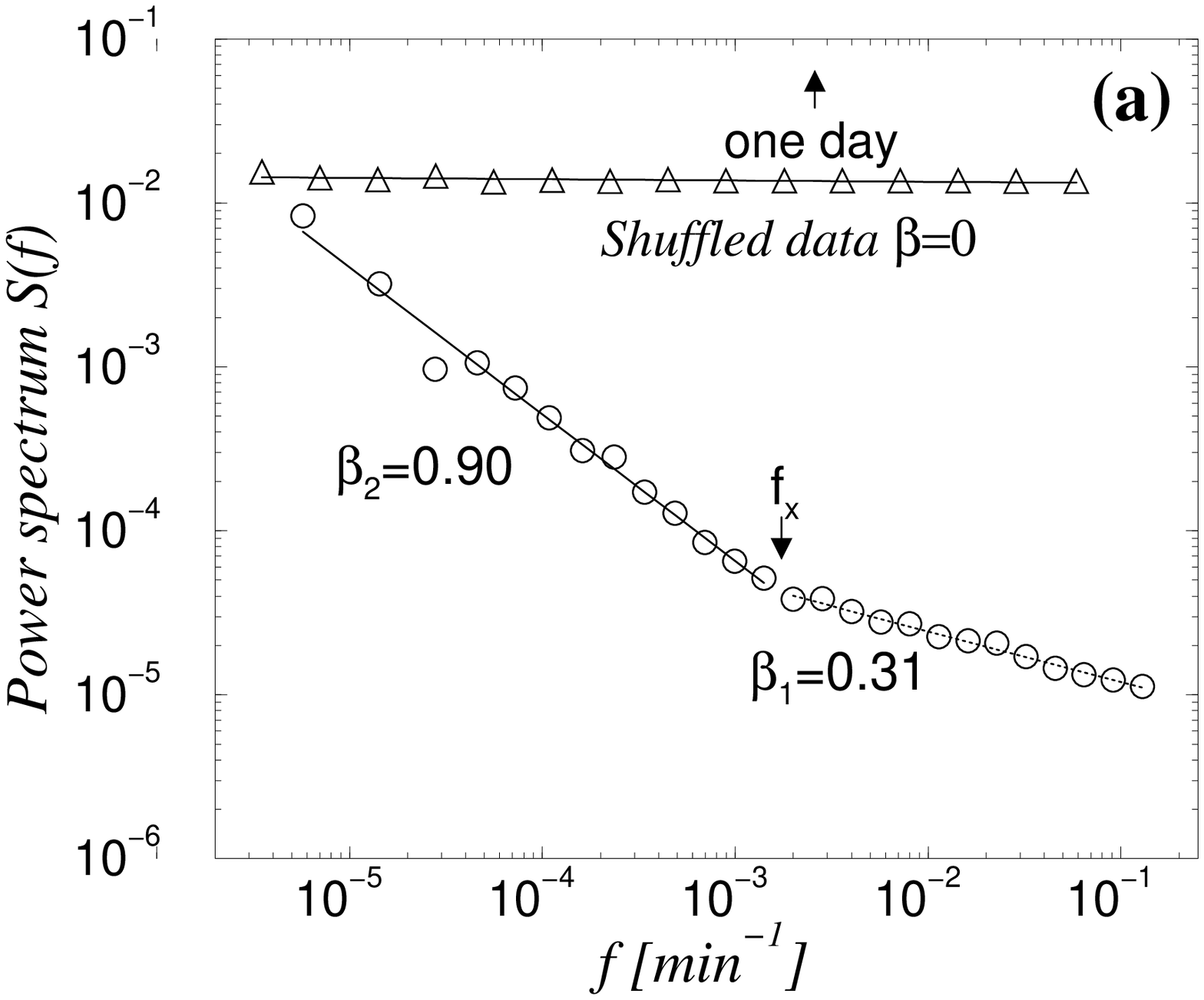}}}
}
\vspace{1cm}
\centerline{
\epsfysize=0.4\columnwidth{{\epsfbox{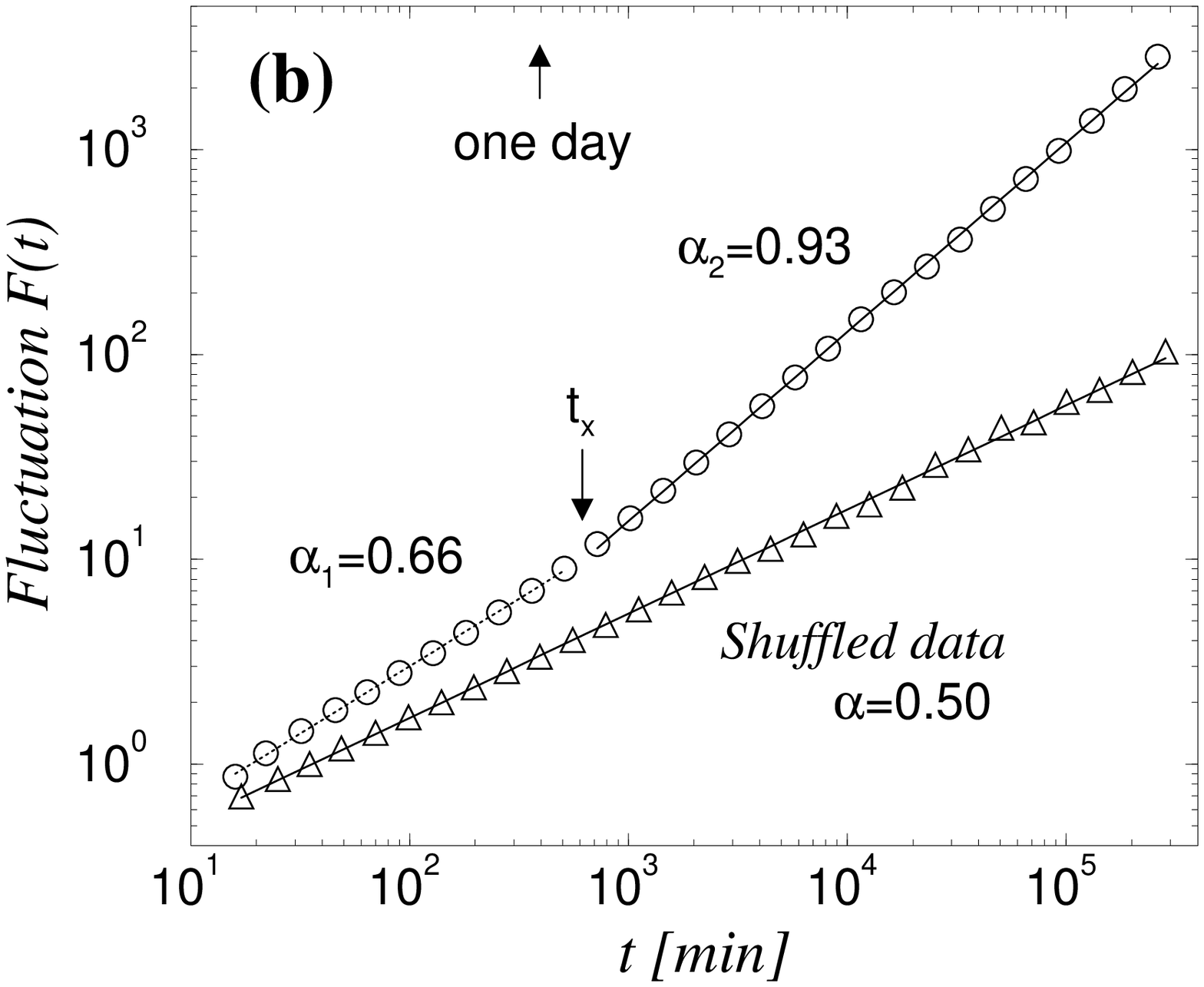}}}
}
\caption{ Plot of (a) the power spectrum $S(f)$ and (b) the detrended
fluctuation analysis $F(t)$ of the absolute values of detrended
increments $g(t)$ with the sampling time interval $\Delta t=1\,$min.
The lines show the best power law fits ($R$ values are better than
$0.99$) above and below the crossover frequency of
$f_\times=(1/570)\,$min$^{-1}$ in (a) and of the crossover time, 
$t_\times=600\,$min in (b). The triangles show the power spectrum and
DFA results for the ``control'', shuffled data (see text for details).
}
\label{powerdfa}
\end{figure}

More accurate results are obtained by the power spectrum
(Fig.~\ref{powerdfa}(a)), which shows that the data fit not one but
rather two separate power laws: for $f>f_\times$, $S(f) \sim
f^{-\beta_1}$, while for $f< f_\times$, $S(f) \sim f^{-\beta_2}$, where
\begin{eqnarray}
\beta_1 = 0.31 \pm 0.02 & \hspace{.5cm}\mbox{ $f > f_\times$} \\
\beta_2 = 0.90 \pm 0.04 & \hspace{.5cm}\mbox{ $f < f_\times$}
\end{eqnarray}
and 
\begin{equation}
f_\times = \frac{1}{570}\,{\rm min^{-1}}\;,
\end{equation}
where $f_\times$ is the crossover frequency.

The DFA method confirms our power spectrum results
(Fig.~\ref{powerdfa}(a)). From the behavior of the power spectrum, we
expect that the DFA method will also predict two distinct regions of
power law behavior, $F(t) \sim t^{\alpha_1}$ for $t < t_\times$ with
exponent $\alpha_1 = 0.66$ and $F(t) \sim t^{\alpha_2}$ for $t >
t_\times$ with $\alpha_2 = 0.95$, where the constant time scale
$t_\times \equiv 1/f_\times$, where we have used the
relation~\cite{Beran94},
\begin{equation}
\alpha=(1+\beta)/2\;.
\end{equation}
Fig.~\ref{powerdfa}(b) shows the results of the DFA analysis. We
observe two power law regions, characterized by exponents,
\begin{eqnarray}
\alpha_1 = 0.66 \pm 0.01 & \hspace{0.5cm} \mbox{ $t < t_\times$} \\
\alpha_2 = 0.93 \pm 0.02 & \hspace{0.5cm} \mbox{ $t > t_\times$}
\end{eqnarray}
in good agreement with the estimates of the exponents from the power
spectrum. The crossover time is close to the result obtained from the
power spectrum, with
\begin{equation}
t_\times\approx1/f_\times\approx 600\,{\rm min}\, 
\end{equation}
or approximately 1.5 trading days.

\subsection{Volatility correlations for individual companies}

The observed correlations in the price changes and the absolute price
changes for the S\&P 500 index raises the question if similar
correlations are present for individual companies which comprise the
S\&P 500 index~\cite{spdiff}.

In the absolute price changes of the individual companies, there is also
a strongly marked intra-day pattern, similar to that of the S\&P 500
index. We compute the intra-day pattern for single companies in the same
sense as before,
\begin{equation}
A_i(t_{day}) \equiv \frac{\sum_{j=1}^{N}|G^j_i(t_{day})|}{N}\;,
\label{eq:ait}
\end{equation}
where time $t_{day}$ refers to the time in the day, the index $i$
denotes companies, and the index $j$ runs over all days---504 days.  In
Fig.~\ref{taq_intraday} we show the intra-day pattern, averaged over all
the 500 companies and contrast it with that of the S\&P 500 stock index.

In order to avoid the intra-day pattern in our quantification of the
correlations, we define a normalized price change for each company,
\begin{equation}
g_i(t)\equiv G_i(t_{day})/A_i(t_{day})\;.
\label{eq:norgit}
\end{equation}

\begin{figure}
\centerline{
\epsfysize=0.45\columnwidth{{\epsfbox{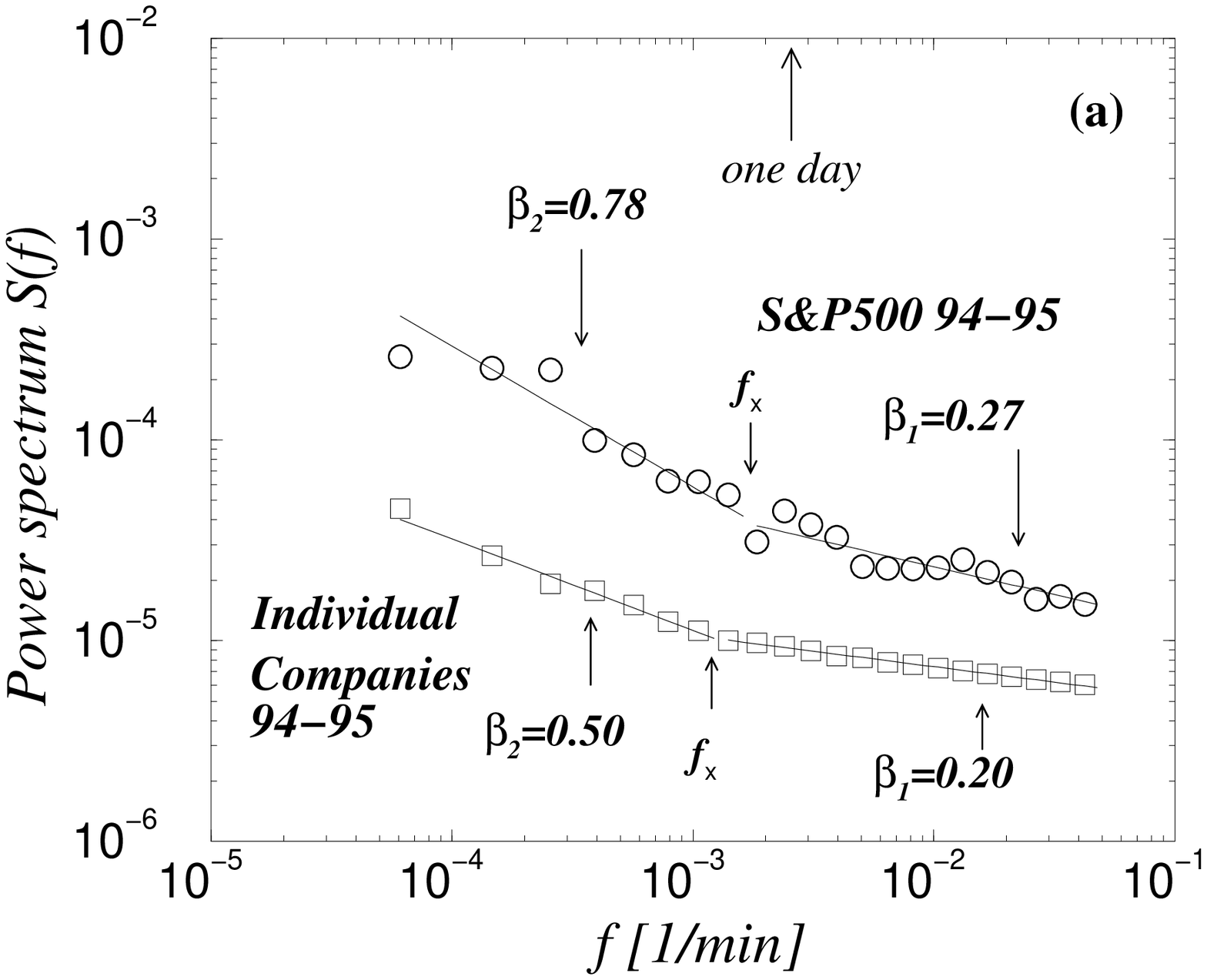}}} 
}
\vspace{0.5cm}
\centerline{
\epsfysize=0.5\columnwidth{{\epsfbox{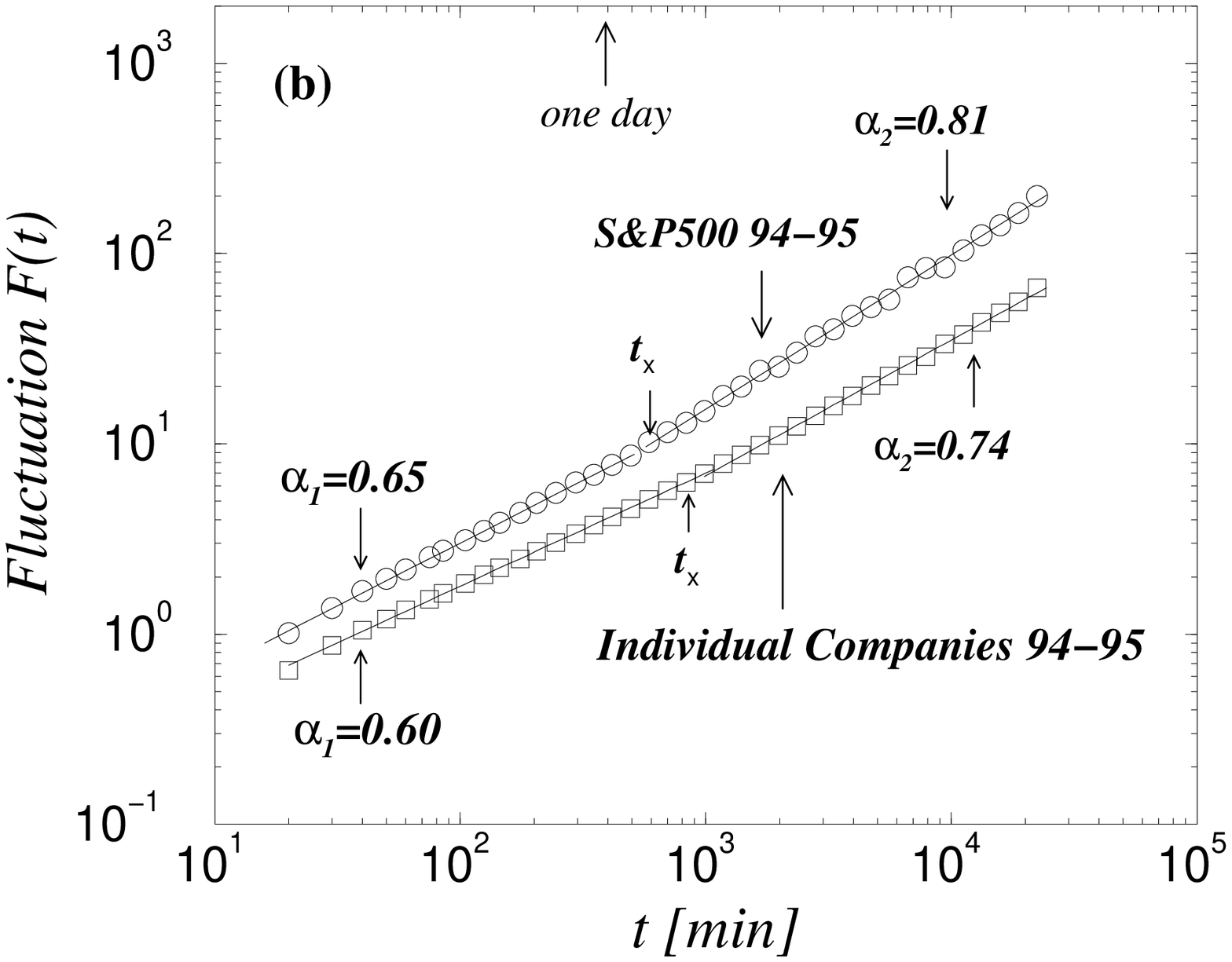}}}
}
\caption{(a) The power spectrum for the absolute values of the
normalized price changes for individual companies, with the sampling
time interval $\Delta t=5\,$min. This is obtained by averaging the power
spectrum $S_i(f)$ for all the 500 companies. We contrast this with the
power spectrum of the S\&P 500 for the same 2-year period
1994-95. Similar to the S\&P 500, we observe two power laws separated by
a crossover frequency. Power law regression fits yield exponents
$\beta_1=0.20$ for the high frequency region and $\beta_2=0.50$ for the
low frequency region. The crossover occurs at approximately 
700$\,$min---slightly larger than that found for the S\&P 500 index. (b)
The average DFA results of $5\,$min sampled $|g(t)|$ for the single
companies, averaged over all 500 companies. It is contrasted with the
result of the S\&P 500 index.  There are two regions characterized by
power laws with exponents $\alpha_1=0.60$ for small time lags and
$\alpha_2=0.74$ for large time lags.}
\label{psd_taq}
\end{figure}

The average autocorrelation function of $g_i(t)$, $i=1,2, \dots 500$,
shows weak correlations up to 10$\,$min, after which there is no
statistically significant correlation. The average autocorrelation
function for the absolute price changes shows long persistence.  We
quantify the long-range correlations by two methods---power spectrum and
DFA. In Fig.~\ref{psd_taq}(a), we show the power spectral density for
the absolute price changes for individual companies and contrast it with
the S\&P 500 index for the same 2-year period. We also observe a similar
crossover phenomena as that observed for the S\&P 500 index. The
exponents of the two observed power laws are,
\begin{eqnarray}
\beta_1 = 0.20 \pm 0.02 & \hspace{0.5cm} \mbox{ $f > f_\times$} \\
\beta_2 = 0.50 \pm 0.05 & \hspace{0.5cm} \mbox{ $f < f_\times$}
\end{eqnarray}
where the crossover frequency is
\begin{equation}
f_\times = \frac{1}{700}\,{\rm min^{-1}} \;.
\end{equation}

In Fig.~\ref{psd_taq}(b), we confirm the power spectrum results by the
DFA method. We observe two power law regimes with 
\begin{eqnarray}
\alpha_1 = 0.60 &\pm & 0.01 \hspace{0.5cm} \mbox{ $t < t_\times$} \\
\alpha_2 = 0.74 &\pm & 0.03 \hspace{0.5cm} \mbox{ $t > t_\times$}
\end{eqnarray}
with a crossover
\begin{equation}
t_\times\approx1/f_\times \approx 700\,{\rm min}\;. 
\end{equation}

The exponents characterizing the correlations in the absolute price
changes for individual companies are on average smaller than what is
observed for the S\&P 500 price changes. This might be due to the
cross-dependencies between price changes of different companies. A
systematic study of the cross-correlations and dependencies will be the
subject of future work~\cite{plerou99}.

\subsection{Additional remarks on power-law volatility correlations} 

Even though several different methods give consistent results, the
power-law correlation of the volatility needs to be tested. It is known
that the power-law correlation could be caused by some artifacts,
e.g. anomaly of the data or the peculiar shape of the distribution etc.

\subsubsection{Data shuffling}

Since we find the volatility to be power-law distributed at the tail, to
test that the power-law correlation is not a spurious artifact of the
long-tailed probability distribution, we shuffled each point of the
$|g(t)|$ randomly for the S\&P500 data. The shuffling operation keeps
the distribution of $|g(t)|$ unchanged, but destroys the correlations in
the time series totally if there are any. DFA measurement of this
randomly shuffled data does not show any correlations and gives exponent
$\alpha = 0.50$ (Fig.~\ref{powerdfa})---confirming that the observed
long-range correlation is not due to the heavy-tailed distribution of
the volatility.

\subsubsection{Outliers removal}

\begin{figure}
\centerline{
\epsfysize=0.4\columnwidth{{\epsfbox{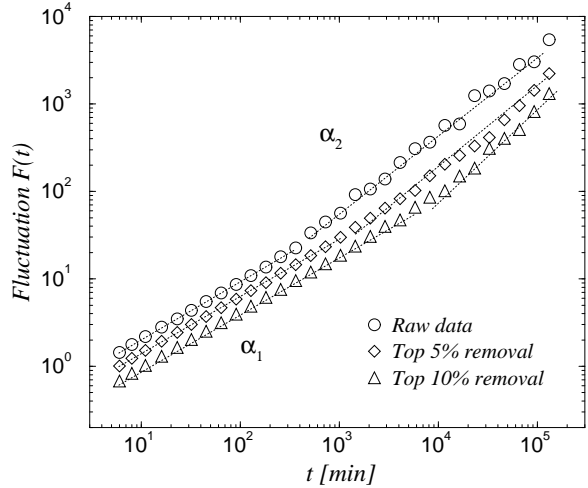}}}
}
\caption{ DFA results of removing top $5\%$ and $10\%$ data points of
the $|g(t)|$ for the S\&P 500 data. The crossover time is approximately
$600\,$min, $1,000\,$min, and $10,000\,$min for the data removing the
top $5\%$ and the top $10\%$ respectively. The DFA exponent $\alpha_1$
for the short time scale does not change, the power law regression fit
gives $\alpha_1 \approx 0.66 $ for all three curves. Regression fits for
the exponent $\alpha_2$ give $0.91 \pm 0.02$, $0.91 \pm 0.03$ and $1.02
\pm 0.04$ for three cases, respectively. }
\label{outlier}
\end{figure}

As an additional test, we study how the outliers (big events) of the
time series $|g(t)|$ affect the observed power-law correlation. We
removed the largest $5\%$ and $10\%$ events of the $|g(t)|$ series and
applied the DFA method to them respectively, the results are shown in
Fig.~\ref{outlier}. Removing the outliers does not change the power-law
correlations for the short time scale. However, that the outliers do
have an effect on the long time scale correlations, the crossover time
is also affected.

\subsubsection{Subregion correlation }

\begin{figure}
\centerline{
\epsfysize=0.5\columnwidth{{\epsfbox{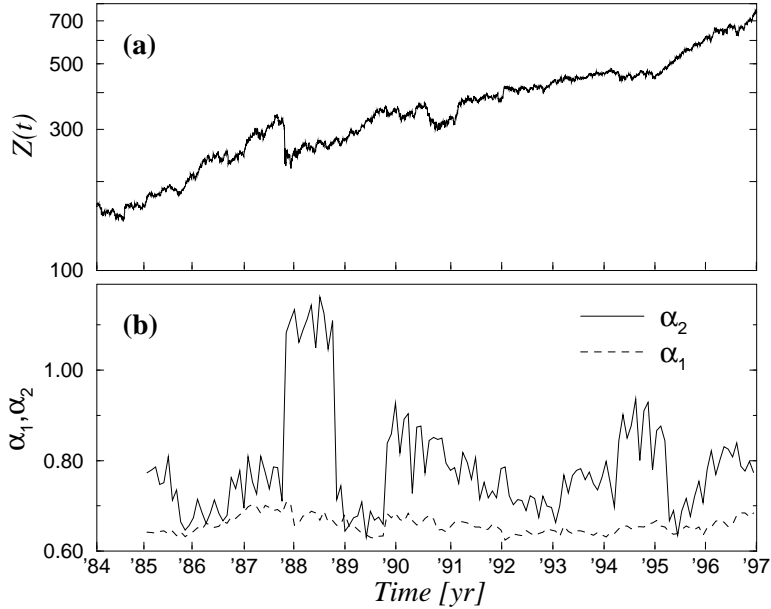}}}
}
\caption{ (a) The S\&P 500 index $Z(t)$ for the 13-year period. (b)
Results of dragging a window of size 1$\,$y down the same data base, one
month at a time, and calculating the best fit exponent $\alpha_1$
(dashed line) and $\alpha_2$ (full line) for the time intervals $t<
t_\times$ and $t>t_\times$ respectively, where $t_\times = 600\,$min.}
\label{alpha}
\end{figure}

The long range correlation and the crossover behavior observed for the
S\&P500 index are for the entire 13-year period. Next, we study whether
the exponents characterizing the power-law correlation are stable,
i.e. does it still hold for periods smaller than 13 years. We choose a
sliding window (with size 1$\,$y) and calculate both exponents
$\alpha_1$ and $\alpha_2$ within this window as the window is dragged
down the data set with one month steps.  We find (Fig.~\ref{alpha}(b))
that the value of $\alpha_1$ is very ``stable'' (independent of the
position of the window), fluctuating slightly around the mean value
2/3. However, the variation of $\alpha_2$ is much greater, showing
sudden jumps when very volatile periods enter or leave the time
window. Note that the error in estimating $\alpha_2$ is also large.

\section{Conclusion}

In this study, we find that the probability density function of the
volatility for the S\&P 500 index seems to be well fit by a log normal
distribution in the center part. However, the tail of the distribution
is better described by a power law, with exponent $1+ \mu \approx 4$,
well outside the stable L\'evy range. The power law distribution at the
tail is confirmed by the study of the volatility distribution of
individual companies, for which we find approximately the same exponent.
We also find that the distribution of the volatility scales for a range
of time intervals. 

We use the Detrended fluctuation analysis and the power spectrum to
quantify correlations in the volatility of the S\&P 500 index and
individual company stocks. We find that the volatility is long-range
correlated. Both the power spectrum and the DFA methods show two regions
characterized by different power law behaviors with a cross-over at
approximately 1.5~days. Moreover, the correlations show power-law decay,
often observed in numerous phenomena that have a self-similar or
``fractal'' origin. The scaling property of the volatility distribution,
its power-law asymptotic behavior, and the long-range volatility
correlations suggest that volatility correlations might be one possible
explanation for the observed scaling behavior~\cite{Mantegna95} for the
distribution of price changes~\cite{Gopi99}.

\section*{Acknowledgments}

We thank L.~A.~N.~Amaral, X.~Gabaix, S.~Havlin, R.~Mantegna, V.~Plerou,
B.~Rosenow and S.~Zapperi for very helpful discussions through the
course of this work, and DFG, NIH, and NSF for financial support.

\section*{Appendix A: Methods to calculate correlations}

\subsection{Correlation function}

The direct method to study the correlation property is the
autocorrelation function,
\begin{equation}
C(t)\equiv\frac{\langle g(t_0)g(t_0+t) \rangle-\langle
g(t_0)\rangle^2}{\langle g^{2}(t_0) \rangle - \langle g(t_0)\rangle^2}\;,
\label{eq:cort}
\end{equation}
where $t$ is the time lag. Potential difficulties of the correlation
function estimation are the following: ({\it i}) The correlation
function assumes stationarity of the time series. This criterion is not
usually satisfied by real-world data.  ({\it ii}) The correlation
function is sensitive to the true average value, $\langle
g(t_0)\rangle$, of the time series, which is difficult to calculate
reliably in many cases. Thus the correlation function can sometimes
provide only qualitative estimation~\cite{Beran94}.

\subsection{Power spectrum}
A second widely used method for calculating correlation properties is the
power spectrum analysis. Note that the power spectrum analysis can only
be applied to linear and stationary (or strictly periodic) time series.

\subsection{Detrended fluctuation analysis}

The third method we use to quantify the correlation properties is called
{\it detrended fluctuation analysis} (DFA)~\cite{peng94,peng95}. The DFA
method is based on the idea that a correlated time series can be mapped
to a self-similar process by integration~\cite{Beran94,peng94,peng95}.
Therefore, measuring the self-similar feature can indirectly tell us
information about the correlation properties.  The advantages of DFA
over conventional methods (e.g. spectral analysis and Hurst analysis)
are that it permits the detection of long-range correlations embedded in
a non-stationary time series, and also avoids the spurious detection of
apparent long-range correlations that are an artifact of
non-stationarities. This method has been validated on control time
series that consist of long-range correlations with the superposition of
a non-stationary external trend~\cite{peng94}. The DFA method has also
been successfully applied to detect long-range correlations in highly
complex heart beat time series~\cite{peng95,iyengan97}, and other
physiological signals~\cite{jeff96pre}.

\begin{figure}
\centerline{
\epsfysize=0.7\columnwidth{{\epsfbox{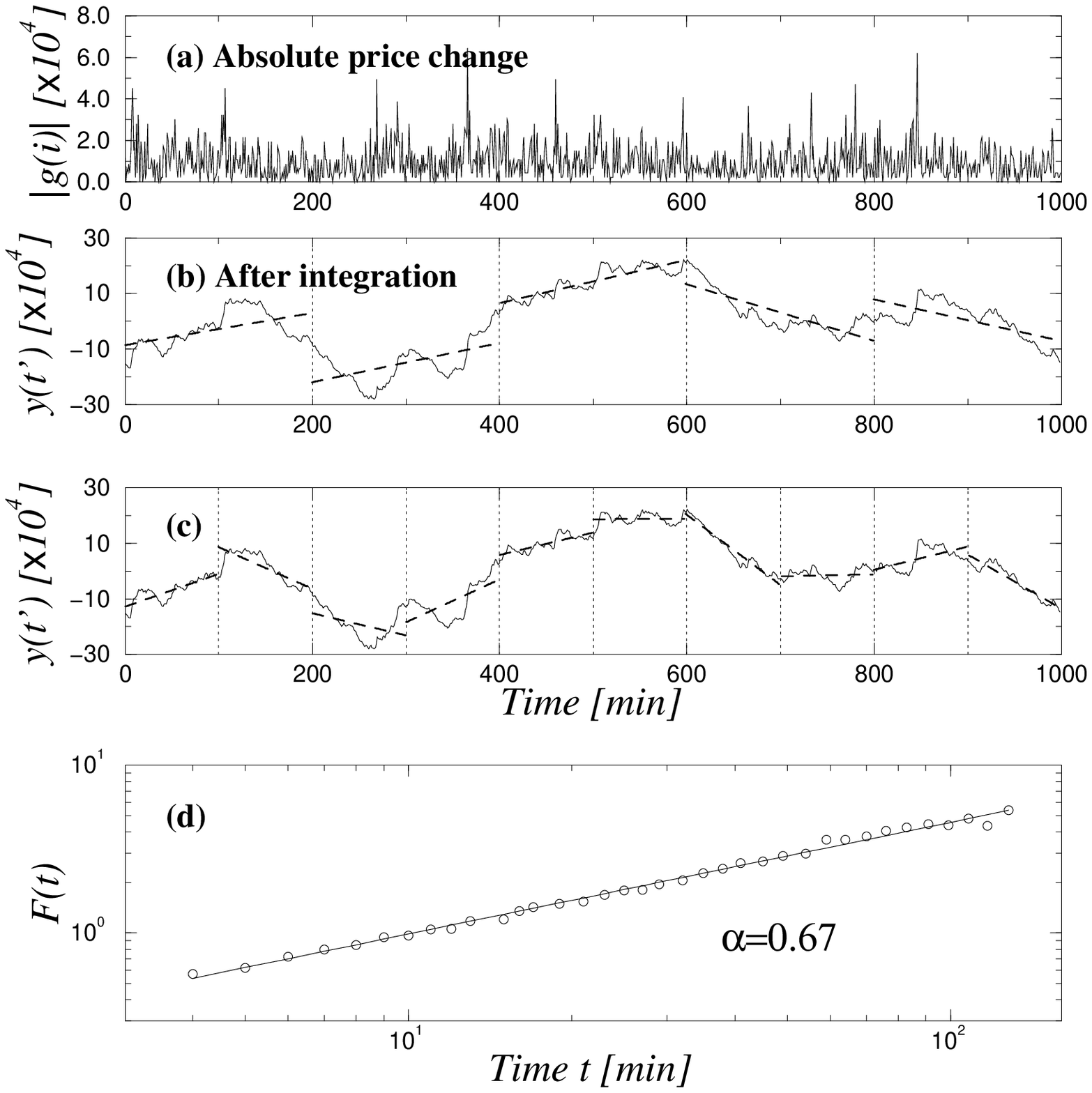}}}
}
\caption{Description of DFA method (see text).}
\label{dfaexplain}
\end{figure}

A description of the DFA algorithm in the context of heart beat analysis
appears elsewhere~\cite{peng94,peng95}. For our problem, we first
integrate $|g(i)|$ time series (with $N$ total data points,
Fig.~\ref{dfaexplain}),
\begin{equation}
y(t')\equiv\sum_{i=1}^{t'}|g(i)|\;.  
\end{equation}
Next the integrated time series is divided into boxes of equal length
$t$. In each box, a least squares line is fit to the data
(representing the {\it trend\/} in that box).  The $y$ coordinate of the
straight line segments is denoted by $y_t(t')$.  Next we de-trend the
integrated time series, $y(t')$, by subtracting the local trend,
$y_t(t')$, in each box. The root-mean-square fluctuation of this
integrated and detrended time series is calculated

\begin{equation}
F(t) =\sqrt{{1\over N}\sum_{t'=1}^N[y(t')-y_t(t')]^2}.
\label{eq:dfa}
\end{equation}

This computation is repeated over all time scales (box sizes) to provide
a relationship between $F(t)$, the average fluctuation, and the box size
$t$. In our case, the box size $t$ ranged from $10\,$min to
$10^{5}\,$min (the upper bound of $t$ is determined by the actual data
length). Typically, $F(t)$ will increase with box size $t$
(Fig.~\ref{dfaexplain}(b)). A linear relationship on a double log graph
indicates the presence of power law scaling. Under such conditions, the
fluctuations can be characterized by a scaling exponent $\alpha$, the
slope of the line relating $\log F(t)$ to $\log t$
(Fig.~\ref{dfaexplain}(b)).

In summary, we have the following relationships between above three methods:

\begin{itemize}
 
\item
For white noise, where the value at one instant is completely
uncorrelated with any previous values, the integrated value, $y(t')$,
corresponds to a random walk and therefore $\alpha = 0.5$, as expected
from the central limit theorem~\cite{feder88,montroll84,Bunde96}. The
autocorrelation function, $C(t)$, is 0 for any $t$ (time-lag) not equal
to 0. The power spectrum is flat in this case.
 
\item
Many natural phenomena are characterized by short-term correlations with
a characteristic time scale, $\tau$, and an autocorrelation function,
$C(t)$ that decays exponentially, [i.e., $C(t) \sim \exp (-t /
\tau)$]. The initial slope of $\log F(t)$ vs.~$\log t$ may be different
from $0.5$, nonetheless the asymptotic behavior for large window sizes
$t$ with $\alpha = 0.5$ would be unchanged from the purely random
case. The power spectrum in this case will show a crossover from $1/f^2$
at high frequencies to a constant value (white) at low frequencies.
 
\item
An $\alpha$ greater than $0.5$ and less than or equal to $1.0$ indicates
persistent long-range power-law correlations, i.e., $C(t)\sim
t^{-\gamma}$. The relation between $\alpha$ and $\gamma$ is
\begin{equation}
\gamma=2-2\alpha\;. 
\end{equation}
Note also that the power spectrum, $S(f)$, of the
original signal is also of a power-law form, i.e., $S(f)\sim
1/f^{\beta}$.  Because the power spectrum density is simply the Fourier
transform of the autocorrelation function, $\beta=1-\gamma=2\alpha-1$.
The case of $\alpha=1$ is a special one which has long interested physicists
and biologists---it corresponds to $1/f$ noise
($\beta=1$).
 
\item
When $0<\alpha<0.5$, power-law {\it anti-correlations\/} are present
such that large values are more likely to be followed by small values
and vice versa~\cite{Beran94}.
 
\item
When $\alpha>1$, correlations exist but cease to be of a power-law form.
 
\end{itemize}
 
The $\alpha$ exponent can also be viewed as an indicator of the
``roughness'' of the original time series: the larger the value of
$\alpha$, the smoother the time series. In this context, $1/f$ noise can
be interpreted as a compromise or ``trade-off'' between the complete
unpredictability of white noise (very rough ``landscape'') and the much
smoother landscape of Brownian noise~\cite{wh78}.

\vspace{1cm}

\newpage

\newpage

\vspace{1cm}

\newpage

\vspace{0.5cm}

\newpage

\vspace{1cm}

\newpage

\newpage

\newpage

\begin{thebibliography}{[199]}
\bibitem{spdiff} 
The majority of the chosen 500 companies belong to the S\&P 500
index. However the companies comprising the S\&P 500 index varies by a
small fraction every year, but this effect is not considerable for the
2-year period.
\end{thebibliography}
\end{document}